\documentclass[prb,twocolumn,showpacs,amsfonts,amssymb,amsmath,superscriptaddress]{revtex4}

	\usepackage{bm}

 	\usepackage{graphicx}

	\newcommand{\vect}[1]{\boldsymbol{#1}}
	\newcommand{\op}[1]{\hat{\boldsymbol{#1}}}
	\newcommand{\hbn}{{\it h}-BN}
	\newcommand{\hbnmath}{\text{{\it h}-BN}}

\begin{document}

\title{Infrared absorption of closely-aligned heterostructures of monolayer and bilayer graphene with hexagonal boron nitride}
\author{D.~S.~L.~Abergel}
\affiliation{Nordita, KTH Royal Institute of Technology and Stockholm University, Roslagstullsbacken 23, SE-106 91 Stockholm, Sweden}
\affiliation{Center for Quantum Materials, KTH and Nordita,
Roslagstullsbacken 11, SE-106 91 Stockholm, Sweden}
\author{Marcin Mucha-Kruczy\'{n}ski}
\affiliation{Department of Physics, University of Bath, Claverton Down, BA2~3FL, United Kingdom}

\begin{abstract}
We model optical absorption of monolayer and bilayer graphene on hexagonal boron nitride for the case of closely-aligned crystal lattices. We show that perturbations with different spatial symmetry can lead to similar absorption spectra. We suggest that a study of the absorption spectra as a function of the doping for almost completely full first miniband is necessary to extract meaningful information about the moir\'{e} characteristics from optical absorption measurements and to distinguish between various theoretical proposals for the physically realistic interaction. Also, for bilayer graphene, the ability to compare spectra for the opposite signs of electric-field-induced interlayer asymmetry might provide additional information about the moir\'{e} parameters.  
\end{abstract}


\maketitle

Graphene and hexagonal boron nitride ({\hbn}) are two representatives of a new class of materials, the two-dimensional atomic crystals.\cite{novoselov_pnas_2005, geim_nature_2013} Both are made of atoms arranged in a single layer of hexagons. However, while graphene, made of carbon, is a gapless semiconductor with a linear dispersion relation for electrons in the vicinity of the two inequivalent corners of the hexagonal Brillouin zone,\cite{abergel_advphys_2010} {\hbn} is an insulator with a band gap of $~6$eV.\cite{watanabe_natmater_2004} It is now possible to place graphene flakes on {\hbn} substrates \cite{dean_natnano_2010} or build stacks with alternating graphene and {\hbn} crystals.\cite{haigh_natmater_2012} Because of the small difference in lattice constants, heterostructures with closely-aligned crystal structures of the two materials, form a moir\'{e} pattern in STM images \cite{xue_natmater_2011} and the dispersion of graphene electrons is reconstructed into minibands as a result of the periodic moir\'{e} perturbation.\cite{yankowitz_natphys_2012, ponomarenko_nature_2013, dean_nature_2013, kindermann_prb_2012, wallbank_prb_2013, wallbank_adp_2015}

Despite considerable theoretical \cite{sachs_prb_2011, song_prl_2013, bokdam_prb_2014, jung_prb_2014, jung_natcomm_2015, dasilva_prb_2015, wijk_prl_2014, moon_prb_2014a, san-jose_prb_2014} and experimental \cite{woods_natphys_2014, shi_natphys_2014} effort, little is known about the actual van der Waals interaction between graphene and {\hbn}. While considering only the inversion-symmetric scalar potential was enough to explain the qualitative features of the initial STM measurements,\cite{yankowitz_natphys_2012} a more complex model involving three inversion-symmetric terms \cite{wallbank_prb_2013} was
used to model some of the transport and capacitance measurements.\cite{ponomarenko_nature_2013, yu_natphys_2014} The same approach was used to demonstrate the importance of the lattice-mixing perturbation term to explain optical absorption measurements.\cite{shi_natphys_2014} However, recent observation of topological currents in graphene/{\hbn} heterostructure \cite{gorbachev_science_2014} means that the moir\'{e} perturbation contains a significant inversion-asymmetric mass term \cite{song_arxiv_2014} while recent theoretical papers suggest that the inversion-asymmetric part of the moir\'{e} perturbation might be comparable to the inversion-symmetric part.\cite{bokdam_prb_2014, jung_natcomm_2015, wijk_prl_2014, moon_prb_2014a, san-jose_prb_2014} In this article, we use four models \cite{wallbank_prb_2013, jung_prb_2014, san-jose_prb_2014, wallbank_adp_2015} developed to describe the effect of the {\hbn} substrate on graphene electrons to investigate the infrared optical absorption of monolayer (MLG) and bilayer (BLG) graphene on {\hbn}. We show that physically different moir\'{e} perturbations can lead to similar absorption spectra, limiting the amount of information about the moir\'{e} perturbation that can be extracted from the spectrum. However, we suggest that careful examination of the absorption spectra for doping in the vicinity of four holes per moir\'{e} unit cell will provide insight into the character of the moir\'{e} perturbation and an experimental technique for distinguishing between the various theoretical proposals for the moir\'e perturbation. Moreover, we show that for BLG on {\hbn} (where, in contrast to MLG on \hbn, the optical absorption has not been investigated previously \cite{shi_natphys_2014, abergel_njp_2013}), it is important to consider the trigonal warping of the electronic band structure when interpreting the absorption spectra, as it leads to shifts of the miniband edges and hence of some of the spectral features. We also discuss the BLG/{\hbn} absorption spectra in the presence of an interlayer asymmetry due to a perpendicular electric field.

\begin{figure}[b]
\centering
\includegraphics[width=1.0\columnwidth]{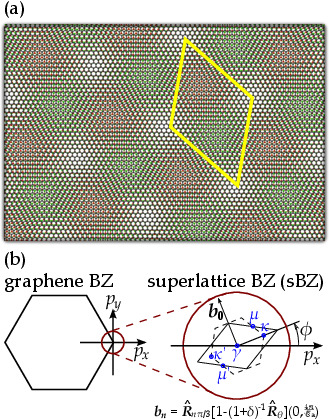}
\caption{(a) Example of the graphene/{\hbn} moir\'{e} superlattice (the lattice mismatch has been enlarged for visual purposes). The black dots denote the carbon atoms in graphene, whereas the red and green correspond to the boron and nitrogen in {\hbn}. The yellow rhombus depicts the superlattice unit cell. (b) Hexagonal Brillouin zone (BZ) of monolayer and bilayer graphene with the valley coordinate system $p_{x},p_{y}$ together with an enlarged view of the supercell Brillouin zone (sBZ) and its own symmetry points $\kappa$ and $\mu$ .}
\label{fig:moire_and_sBZ}
\end{figure}

For a MLG/BLG crystal with a lattice constant $a_{\mathrm{G}}=2.46$\AA{} placed on top of a {\hbn} flake with a lattice constant $a_{\hbnmath}=(1+\delta)a_{\mathrm{G}}$, the lattice mismatch $\delta=1.8\%$,\cite{xue_natmater_2011} together with a possible misalignment of the two lattices given by the angle $\theta$, create for small $\theta\lesssim 2^{\circ}$ a periodic structure shown schematically in Fig.~1(a). This planar superlattice can be described using a set of reciprocal-lattice vectors $\vect{b}_{n}\!\!=\!\!\op{R}_{n\pi/3}\!\left[1\!-\!(1\!+\!\delta)^{-1}\op{R}_{\theta}\right]\!(0,\frac{4\pi}{\sqrt{3}a})$, $n\!=\!0,1,\dots,5$, where
$\op{R}_{\varphi}$ stands for anticlockwise rotation by angle $\varphi$ and $b\!=\!\!|\vect{b}_{n}|\!\!\approx\!\!\frac{4\pi}{\sqrt{3}a}\!\sqrt{\delta^{2}\!+\!\theta^{2}}$.\cite{wallbank_prb_2013} Note that this set both rotates by $\phi(\theta)$ and changes its size as a function of $\theta$, see Fig.~\ref{fig:moire_and_sBZ}(b), which is especially important when considering the zone-folding of the trigonally warped band structure of BLG.\cite{mucha-kruczynski_prb_2013}

In order to obtain the perturbed band structure and wave functions, and hence to compute the absorption spectra, we use a recently proposed phenomenological model constructed to describe the perturbation caused by the closely-aligned {\hbn} substrate on graphene electrons.\cite{wallbank_prb_2013} Within this description, all symmetry-allowed terms containing the dominant first harmonics of the moir\'{e} perturbation are added to the effective Dirac-like Hamiltonian describing electrons in MLG in the vicinity of one of the two corners (valleys) $K$/$K'$ of the hexagonal Brillouin zone. This model has been also applied to the BLG/{\hbn} heterostructure \cite{mucha-kruczynski_prb_2013} by assuming that only the bottom layer is affected by the {\hbn} substrate. The resulting Hamiltonians $\op{H}^{\mathrm{MLG}}$ and $\op{H}^{\mathrm{BLG}}$ for electrons in monolayer and bilayer graphene, respectively, can be written as \cite{wallbank_prb_2013, mucha-kruczynski_prb_2013, wallbank_adp_2015}

\begin{table}[b]
\centering
\caption{Values in meV of parameters from Eq.~\eqref{eqn:hamiltonians} for the four models used in this work to calculate the absorption spectra of MLG/{\hbn} and BLG/{\hbn} heterostructures. The given values all correspond to perfectly aligned ($\theta=0$) structures.}
\label{table}
\resizebox{\columnwidth}{!}{%
\begin{tabular}{|c|r|r|r|r|r|r|r|c|}
\hline
\multicolumn{1}{|l|}{model} & \multicolumn{1}{c|}{$u_{0}^{+}$} & \multicolumn{1}{c|}{$u_{1}^{+}$} & \multicolumn{1}{c|}{$u_{3}^{+}$} & \multicolumn{1}{c|}{$u_{0}^{-}$} & \multicolumn{1}{c|}{$u_{1}^{-}$} & \multicolumn{1}{c|}{$u_{3}^{-}$} & \multicolumn{1}{c|}{$\Delta$} & \multicolumn{1}{c|}{Ref.} \\ \hline
(1)                         & 11                           & -21                          & -18                          & 0                                    & 0                                    & 0                                    & 0                             & \cite{wallbank_prb_2013, wallbank_adp_2015}                       \\ \hline
(2)                         & 0                            & 0                            & 0                            & -11                                  & -21                                  & -18                                  & 0                             & \cite{wallbank_prb_2013, wallbank_adp_2015}                       \\ \hline
(3)                         & 1.26                         & 0.7                          & -0.36                        & 8.98                                 & -7.31                                & -5.63                                & 3.74                          & \cite{jung_prb_2014, dasilva_prb_2015}                       \\ \hline
(4)                         & 2                            & 21                           & -0.06                        & 5.2                                  & -42                                  & -5.9                                 & 5.3                           & \cite{san-jose_prb_2014}                       \\ \hline
\end{tabular}
}
\end{table}

\begin{align}\begin{split}\label{eqn:hamiltonians}
& \op{H}^{\mathrm{MLG}} = v\vect{\sigma}\cdot\vect{p}+\op{\delta\!H}\\
& \op{\delta\!H}= (u^{+}_{0}f_{+}+u^{-}_{0}f_{-}) + \tau_{z}\sigma_{z}(u^{+}_{3}f_{-}+u^{-}_{3}f_{+}) \\ 
& + \tau_{z}\vect{\sigma}\cdot[\vect{l_{z}}\times\nabla(u^{+}_{1}f_{-}+u^{-}_{1}f_{+})] + \Delta\tau_{z}\sigma_{z},\\
& f_{+} = \sum_{n}e^{i\vect{b_{n}}\cdot\vect{r}}, f_{-} = i\sum_{n}(-1)^{n}e^{i\vect{b_{n}}\cdot\vect{r}}, \\
& \op{H}^{\mathrm{BLG}} = \left( \begin{array}{cc} v\vect{\sigma}\cdot\vect{p}+\op{\delta\!H}+u & \op{T} \\ \op{T}^{\dagger} & v\vect{\sigma}\cdot\vect{p}-u \end{array} \right), \\
& \op{T} = \frac{1}{2}\gamma_{1}(\tau_{z}\sigma_{x}\!-\!i\sigma_{y})+\frac{1}{2}v_{3}(\sigma_{x}\!+\!i\tau_{z}\sigma_{y})(p_{x}\!+\!i\tau_{z}p_{y}).
\end{split}\end{align}
The MLG Hamiltonian $\op{H}^{\mathrm{MLG}}$ above is written in the basis of the Bloch states on sublattices $\{\phi(A),\phi(B)\}$ in the $K$ valley and $\{\phi(B),-\phi(A)\}$ in $K'$, whereas for the BLG Hamiltonian $\op{H}^{\mathrm{BLG}}$ the basis is $\{\phi(A),\phi(B),\phi(\tilde{A}),\phi(\tilde{B})\}$ near $K$ and
$\{\phi(B),-\phi(A),\phi(\tilde{B}),-\phi(\tilde{A})\}$ near $K'$. In the BLG case, the tilde denotes the sublattice in the top layer. We also use two sets of Pauli matrices $\sigma_{i}$, $\vect{\sigma}=(\sigma_{x},\sigma_{y})$, and $\tau_{i}$ acting in the sublattice and valley space, respectively. The terms $v\vect{\sigma}\cdot\vect{p}$, $v\approx 10^{6}$m/s,\cite{jiang_prl_2007} arise due to the electron hopping from a carbon atom to one of the three nearest neighbors in the same layer. In MLG, this gives rise to the conical electronic dispersion relation. In BLG, due to the interlayer coupling block $\op{T}$, which includes the direct interlayer coupling $\gamma_{1} = 0.38$eV \cite{kuzmenko_prb_2009} the electronic dispersion is roughly parabolic at energies $\epsilon\ll\gamma_{1}$.\cite{mccann_prl_2006} Also present in $\op{T}$ is the skew interlayer coupling, $v_{3}\approx 0.1v$ \cite{kuzmenko_prb_2009} which induces a trigonal distortion in the low energy band structure. We also include the interlayer asymmetry $u$ which can be induced by applying electric field perpendicular to BLG and which opens a band gap in the electronic spectrum.\cite{mccann_prl_2006, mccann_prb_2006} Finally, the moir\'{e} perturbation $\op{\delta\!H}$ consists of four terms: the first describes a simple potential modulation, the second the local $A$-$B$ sublattice asymmetry due to the substrate, and the third the spatial modulation of hopping between the $A$ and $B$ sublattices. The final term, characterised by the parameter $\Delta$, describes a non-zero average value of the sublattice asymmetry. Within each of the first three contributions to the moir\'{e} perturbation, the first term inside the round bracket, characterised by the dimensionless parameters $u_{i}^{+},\,i=0,1,3$, describes the inversion symmetric part of the perturbation. Correspondingly, the second term in each round bracket, characterised by one of the dimensionless parameters $u_{i}^{-},\,i=0,1,3$, represents the inversion asymmetric part of the perturbation. Note that in the case of BLG, the perturbation enters only in the part of the Hamiltonian describing the bottom layer since, because of the exponential decay of the $2p_{z}$ orbital wave function with increasing distance \cite{house_book_2004}, we assume that the effect of {\hbn} on the top layer of graphene, twice as far from the substrate as the bottom one, can be neglected.\cite{footnote_couplings}

The position-dependent moir\'{e} perturbation terms we wrote in Eq.~\eqref{eqn:hamiltonians} represent all symmetry-allowed terms generated by the first harmonic of the moir\'{e} within the continuum model. Various physical effects, like strain generated due to mutual relaxation of graphene and {\hbn} lattices \cite{woods_natphys_2014} or electron-electron interaction \cite{song_prl_2013}, contribute to the currently unknown values of the parameters $\{u_{i}^{+}$, $u_{i}^{-}\}$ and $\Delta$. Here, we use values as provided by four models published previously. All the sets, summarised in Table \ref{table}, were originally obtained for a MLG/{\hbn} heterostructure with no angular misalignment ($\theta=0$). The first two, \cite{wallbank_prb_2013, wallbank_adp_2015} (1) and (2), result from representing {\hbn} as a lattice of point-charges. Model (1) assumes that the inversion-symmetric part of the perturbation is dominant, whereas for model (2) only the inversion-asymmetric part is non-zero. The strength of the perturbation is then governed by a single parameter, which we set by requiring that $\sum_{i}(|u_{i}^{+}|+|u_{i}^{-}|)=50$meV for both of the models. Parameters for the model (3) \cite{jung_prb_2014, dasilva_prb_2015} have been obtained using DFT. Model (4) \cite{san-jose_prb_2014} borrows the information on the stacking-dependent adhesion energies from DFT and uses them as input in an analytic description of the low-energy band structure taking into account the elastic energy of graphene.\cite{san-jose_prb_2014a} We point out that both models (3) and (4) explicitly include the influence on the electrons of deformations of the graphene lattice arising due to its structural relaxation on top of {\hbn}. Such relaxation minimises the total energy of the crystal through the interplay between lowering local van der Waals interaction energy but increasing the elastic energy by introducing local strains. The evidence of such strains in graphene was found in Ref.~\onlinecite{woods_natphys_2014}. Note that it is known that the dominant effect of deformations in graphene crystals is the generation of a gauge field \cite{neto_rmp_2009} with sublattice and valley structure identical to the third term in $\op{\delta\!H}$ in Eq.~\eqref{eqn:hamiltonians}. Indeed, according to Ref.~\onlinecite{san-jose_prb_2014} and model (4), the inhomogeneous, moir\'{e}-periodic strain in graphene lattice generates significant parameters $u_{1}^{+}$ and $u_{1}^{-}$, see Table \ref{table}. This is however not the case according to model (3). Deformations also induce terms in $\op{\delta\!H}$ with higher harmonics of the moiré. However, these only lead to second order corrections to the energy of the lowest minibands.

\begin{figure}[b]
\centering
\includegraphics[width=1.0\columnwidth]{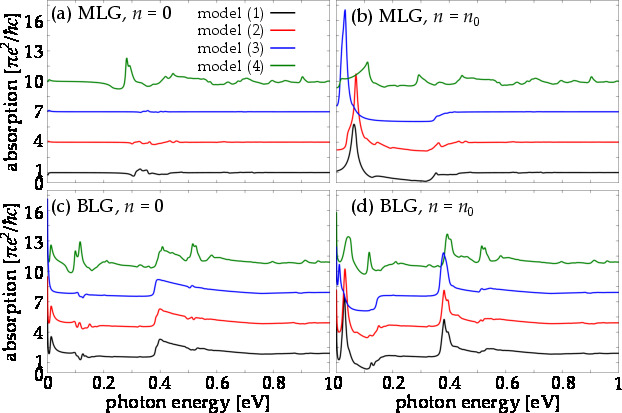}
\caption{Absorption spectra for MLG and BLG on {\hbn} with hole density $n=0$ or $n=n_{0}=\frac{b^{2}\sqrt{3}}{2\pi^{2}}$, equivalent to filling of one separated miniband with holes. For clarity, the consecutive spectra have been shifted by $3\frac{\pi e^{2}}{\hbar c}$ along the $y$ axis.}
\label{fig:DP_and_NP}
\end{figure}

In order to plot the optical absorption spectra shown in this work, we calculate the absorption coefficient $g(\omega)$ for incident light with frequency $\omega$ and polarisation $\vect{e}=(e_{x},e_{y})$,
\begin{align}
g(\omega) = \frac{8\pi\hbar}{c\omega S}\Im\sum_{\vect{p},\lambda,\lambda'}\frac{f_{\vect{p}\lambda'}-f_{\vect{p}\lambda}}{\hbar\omega+\epsilon_{\vect{p}\lambda}-\epsilon_{\vect{p}\lambda'}+i\eta}M_{\alpha\beta}^{\lambda\lambda'}e_{\alpha}^{*}e_{\beta}, \nonumber
\end{align}
where $\alpha,\beta=x,y$, $\epsilon_{\vect{p}\lambda}$ stands for the miniband energy found by diagonalisation of the appropriate Hamiltonian in Eq.~\eqref{eqn:hamiltonians}, $f_{\vect{p}\lambda}$ are the occupation numbers, $S$ is the normalisation area of the miniband plane wave states and $\eta$ is the broadening of the energy states (which, unless stated otherwise, we take to be $\eta=3$meV). The indices $\lambda$ and $\lambda'$ represent the remaining quantum numbers: spin, valley, and miniband index. We also numerically compute the eigenstates of the Hamiltonians in Eq.~\eqref{eqn:hamiltonians}, in order to calculate the matrix elements of the current operator 
\begin{align}
M_{\alpha\beta}^{\lambda\lambda'}=\langle \vect{p}\lambda |
\op{j}_{\alpha}^{*} | \vect{p}\lambda' \rangle \langle \vect{p}\lambda' 
| \op{j}_{\beta} | \vect{p}\lambda \rangle, \nonumber
\end{align}
where $\op{j}_{\alpha}= \partial \hat{\vect{H}}/\partial \vect{p}_\alpha= ev\sigma_{\alpha}$ are the Dirac current operators. In the absence of a moir\'e perturbation, the monolayer absorption is flat with a value of $\pi e^2/\hbar c$,\cite{gusynin_prb_2006} while the bilayer shows a pronounced peak at $\hbar \omega = \gamma_1$.\cite{abergel_prb_2007, nicol_prb_2008}

In Fig.~\ref{fig:DP_and_NP}, we show optical absorption spectra calculated for all the four perturbation models in Table I for both MLG (top row) and BLG (bottom row). For the spectra in the left column, we assume that the structure is neutral (chemical potential $\mu$ is at the Dirac point for MLG and the neutrality point for BLG). The spectra in the right column are for $p$--doped structures with hole density $n_{0}=\tfrac{b^{2}\sqrt{3}}{2\pi^{2}}$ equivalent to four holes per moir\'{e} unit cell ($n_{0} \approx 2.5\times 10^{12} \mathrm{cm}^{-2}$ for $\theta = 0$). If the first valence miniband is isolated from the rest of the spectrum (for example, by a secondary Dirac point somewhere at the edge of the sBZ as suggested by experiments,\cite{ponomarenko_nature_2013, dean_nature_2013}) this hole density corresponds to the full filling of this miniband. As a result, optical transitions at low frequencies are sensitive to the details of the reconstructed miniband spectrum. 

For the neutral structures, one expects the absorption spectra to be modified in the frequency range $\hbar\omega_{\mathrm{MLG}}\approx \hbar vb$ for MLG and $\hbar\omega_{\mathrm{BLG}}\approx\frac{(\hbar vb)^{2}}{\gamma_{1}}$ for BLG, as these energies correspond to optical transitions from the reconstructed electronic states at the boundary of sBZ in the valence band to matching states on the conduction-band side. However, the first three models show only weak modulation of the absorption spectra, both for MLG and BLG. As a result, despite significant differences in the weighting of the parameters $u_{i}^{+}$ and $u_{i}^{-}$ in each of the sets and important differences in the underlying physics (e.g.~presence or lack of inversion symmetry), the presented spectral features do not offer a clear method of distinguishing between the perturbation models. For example, in a recent experimental work,\cite{shi_natphys_2014} the optical conductivity data was fitted using model (1) and the results interpreted as a confirmation of the importance of the sublattice mixing term, in particular $u_{1}^{+}$. However, our results indicate that distinction between the influences of $u_{1}^{+}$ and $u_{1}^{-}$ or, as an alternative example, between models (1) and (2), requires more careful analysis. In fact, potentially very different combinations of the moir\'{e} perturbation parameters might lead to similar spectra for particular choices of chemical potential. However, if model (4) is the physically relevant description, then the presence of strong absorption peaks at $\hbar \omega \approx 0.3\mathrm{eV}$ for the monolayer, and $\hbar \omega \approx 0.1\mathrm{eV}, 0.5\mathrm{eV}$ for the bilayer will be experimentally distinguishable. This is also evidenced in the right column in Fig.~\ref{fig:DP_and_NP} where we model spectra of MLG and BLG structures $p$--doped with hole density $n_{0}$. Again, the optical spectra for models (1) and (2) look very similar, with the most significant feature being a low-energy peak indicating optical transitions between the second and first miniband in the valence band. This peak, although higher and shifted to even lower frequencies, appears also in the spectrum for model (3). Again, model (4) yields the most distinctive spectrum. 

\begin{figure}[b]
\centering
\includegraphics[width=1.0\columnwidth]{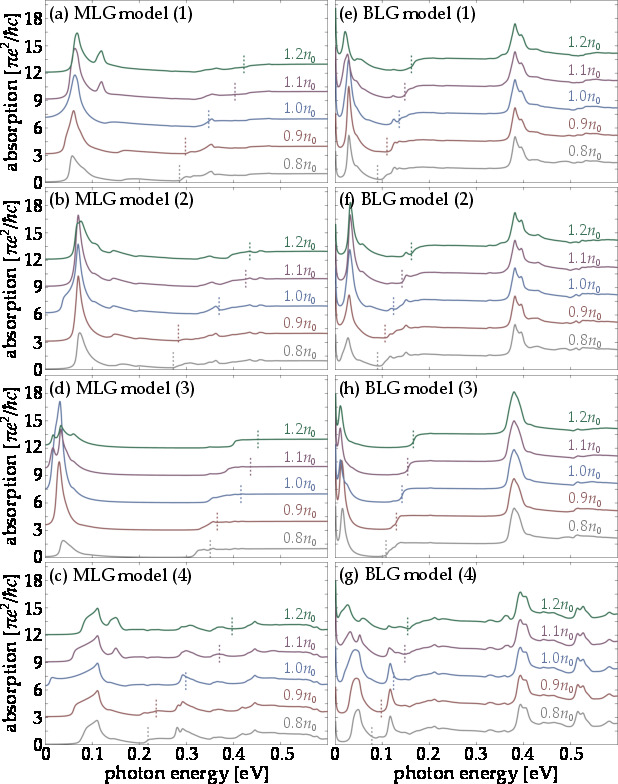}
\caption{Absorption spectra for structures hole-doped to $0.8n_{0}$, $0.9n_{0}$, $n_{0}$, $1.1n_{0}$, and $1.2n_{0}$. For clarity, the consecutive spectra have been shifted by $3\frac{\pi e^{2}}{\hbar c}$ along the $y$ axis. For each spectrum, the dashed vertical line indicates the $2\mu$ threshold for Pauli blocking below which the absorption for an unperturbed MLG/BLG would be zero.}
\label{fig:MLG_BLG_filling}
\end{figure}

We now move beyond a ``single spectrum'' picture, which concentrated on analysing the absorption spectrum for a given carrier density, and discuss changes in optical spectra as a function of increasing hole doping $n\approx n_{0}$. In Fig.~\ref{fig:MLG_BLG_filling}, we show the evolution of optical absorption for the four moir\'{e} perturbation models as the doping is varied from $0.8n_{0}$ to $1.2n_{0}$. For this density range, absorption spectra at low energies are dominated by features due to the moir\'{e} perturbation (for an unperturbed MLG/BLG spectrum with chemical potential $\mu$, optical transitions with energy $\epsilon<2\mu$ are forbidden due to Pauli blocking). For MLG for both models (1) and (2) [panels (a) and (b)], the low-energy peak due to transitions connecting states at the boundary of the sBZ in the first and second valence miniband grows in size with $n$ increasing up to $n\sim 1.1n_{0}$. A shoulder also develops for $n \approx n_0$. However, the spectrum for model (1) displays a clear side feature to the right of the main peak which appears when filling the second miniband which is a clear experimental indicator of strong inversion-symmetric perturbation. For model (3), the peak decreases rapidly after the hole density increases beyond $n_{0}$ and splits into a double-peak feature. In comparison, for model (4), the height of the low-energy peak does not change much as a function of increasing hole density. Its width increases as the first miniband is fully filled and then decreases to roughly the same shape. However, similarly to model (1), a second smaller peak appears at slightly higher frequencies.

For the BLG spectra, we notice a peak at lower frequencies than for monolayer, at $\sim50$meV. For model (1), this peak grows for hole densities lower than $n_{0}$ and then decreases for $n>n_{0}$. This is in contrast to the behaviour for model (2), for which this peak remains of similar height for $n=0.8n_{0}$ and $n=0.9n_{0}$ but increases in the vicinity of $n_{0}$ and remains high with further increase of $n$. This is a second clear way to experimentally distinguish model (1) from model (2). For model (3), the low-energy absorption peak grows for the hole density increasing from $n=0.8n_{0}$ to $n=0.9n_{0}$ but decreases already for $n=n_{0}$ and remains small with further increase of the hole density. The spectrum obtained for model (4) is the only one for BLG that displays more clearly distinguished features. First of all, the low-energy absorption peak does not change much with $n\leq n_{0}$. However, for $n>n_{0}$ this peak splits into two, with the one at higher energy decaying quicker than the one at lower energy as the hole density is increased. Moreover, a second peak above $0.1$eV exists for $n\leq n_{0}$. 

\begin{figure}[t]
\centering
\includegraphics[width=1.0\columnwidth]{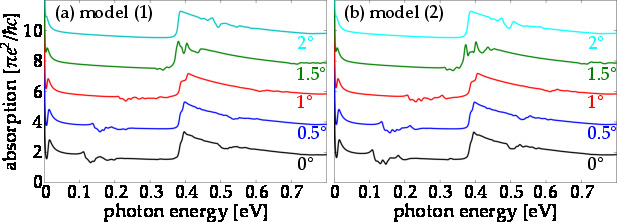}
\caption{Absorption spectra for BLG/{\hbn} structures [models (1) and
(2) only] with various angular misalignments between the crystalline axes of the two crystals. For clarity, the consecutive spectra have been shifted by $2\frac{\pi e^{2}}{\hbar c}$ along the $y$ axis.}
\label{fig:BLG_angle_dependence}
\end{figure}

An interesting open question is the dependence of the electronic moir\'{e} perturbation effects on the misalignment between the crystalline directions of graphene and {\hbn}. Making use of the dependence of $u_{i}^{+}$ and $u_{i}^{-}$ on $\theta$ predicted for models (1) and (2),\cite{wallbank_prb_2013, wallbank_adp_2015} we show in Fig.~\ref{fig:BLG_angle_dependence} the absorption spectra for the BLG/{\hbn} heterostructure for those two models (see Ref.~\onlinecite{abergel_njp_2013} for the MLG/{\hbn} heterostructure) and a range of misalignment angles $\theta$. Because the perturbation parameters decrease with increasing misalignment, the spectral features due to the moir\'{e} weaken as well. However, the shift of the spectral features toward higher frequencies is visible as the characteristic energy of the moir\'{e}, $\hbar vb$, increases with $\theta$. In particular, for misalignment $\theta\sim 1^{\circ}$--$1.5^{\circ}$, the peak at the energy $\sim\gamma_{1}$ due to optical transitions from the low-energy bands to the high-energy split bands \cite{abergel_prb_2007} is modified because at these angles $\frac{vb}{\gamma_{1}}\approx 1$. 

\begin{figure}[b]
\centering
\includegraphics[width=1.0\columnwidth]{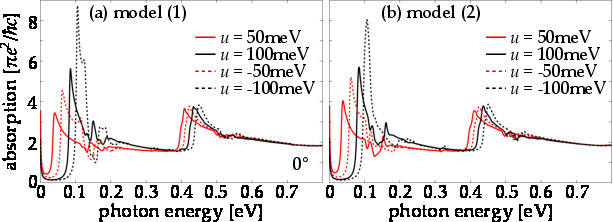}
\caption{Absorption spectra for BLG/{\hbn} heterostructures [models (1) and (2) only] in the presence of the interlayer asymmetry $u$.}
\label{fig:BLG_gaps}
\end{figure}

One of the unique features of BLG is the possibility of opening a bandgap in the electronic spectrum by breaking layer symmetry with external electric field applied perpendicular to the graphene layers.\cite{mccann_prl_2006, mccann_prb_2006} Band gaps $\Delta_{\mathrm{g}}\approx |u| \approx 200$meV have been opened with high electric fields.\cite{zhang_nature_2009, varlet_prl_2014} The interlayer asymmetry $u$ breaks inversion symmetry but preserves electron-hole symmetry \cite{mccann_prb_2006} and hence, in the absence of the moir\'{e} perturbation, absorption in the case of asymmetry $u$ is the same as for $-u$.\cite{nicol_prb_2008} In comparison, a generic graphene--{\hbn} interaction described by the parameters in Eq.~\eqref{eqn:hamiltonians} breaks electron-hole symmetry and opens a gap $\Delta_{0}$ quadratic in the moir\'{e} parameters $u_{i}^{+},u_{i}^{-}$.\cite{mucha-kruczynski_prb_2013} As a result, for both the moir\'{e} perturbation and interlayer asymmetry, the gap at the neutrality point is $\Delta_{\mathrm{g}}\approx|u+\Delta_{0}|$, yielding different gaps for $u$ and $-u$. This is seen in Fig.~\ref{fig:BLG_gaps} where we show optical absorption spectra calculated for BLG/{\hbn} heterostructures with moir\'{e} perturbation effects described by models (1) and (2) in Table I and in the presence of $|u|=50$meV and $|u|=100$meV. At frequencies close to zero, the absorption falls to zero \cite{footnote_on_low_frequency_behaviour}, indicating presence of a gap in the spectrum. This gap is larger for $u<0$ because for both models $\Delta_{0}<0$. Because of that difference in the band gap at the neutrality point, although the general features of the spectra for $u$ and $-u$ are similar, those for $u<0$ are consistently shifted to higher frequencies. We suggest that, if $\Delta_{0}$ is experimentally significant in BLG, then comparing optical absorption spectra for opposite interlayer asymmetries $u$ and $-u$ should reveal the sign of $\Delta_{0}$ and through the dependence of $\Delta_{0}$ on $u_{i}^{+},u_{i}^{-}$ potentially some more information about the moir\'{e} perturbation characteristics.

\begin{figure}[b]
\centering
\includegraphics[width=1.0\columnwidth]{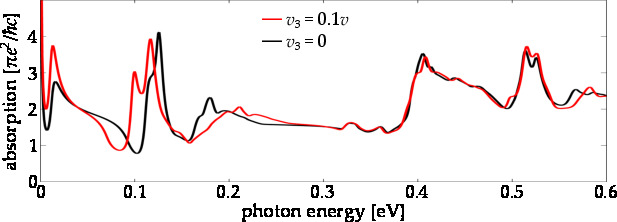}
\caption{Absorption spectra for BLG/{\hbn} heterostructure with moir\'{e} parameters according to model (4) in Table I without (black solid line) and including the effects of trigonal warping.}
\label{fig:trigonal_warping}
\end{figure}

Finally, in Fig.~\ref{fig:trigonal_warping} we present comparison of BLG optical absorption spectra calculated taking into account or neglecting the interlayer skew coupling term parametrised by the velocity $v_{3}$ in Eq.~\eqref{eqn:hamiltonians}. For this example, we used parameters from model (4) and assumed chemical potential to lie between the valence and conduction bands. For $v_{3}=0$, the electronic spectrum has circular symmetry and so all states with the same magnitude of momentum $p=|\vect{p}|$ contribute to the absorption at the same frequency. However, nonzero trigonal warping shifts the energy of an electronic state by amount depending on the polar angle of $\vect{p}$. As a result, both the position and intensity of the spectral features change. For that reason, all spectra in Figs.~\ref{fig:DP_and_NP}-\ref{fig:BLG_gaps} for the BLG/{\hbn} heterostructure were carried out with $v_3 = 0.1v$.

In summary, we have discussed the optical absorption spectra for MLG and BLG on closely aligned {\hbn} using four models developed previously to describe the effect of the substrate on graphene electronic dispersion. We showed that, unfortunately, physically very different moir\'{e} perturbations can lead to similar absorption spectra. A detailed investigation of the spectra for samples with (almost) completely filled first miniband will provide more information about the details of the electronic perturbation due to moir\'{e}. For BLG on {\hbn}, additional information can be obtained by measuring absorption spectra in double-gated devices. However, for this system it is important to include the trigonal warping of the electronic band structure during theoretical analysis.

During the final stages of writing of this paper, we became aware of the work in Ref.~\onlinecite{dasilva_arxiv_2015}, which calculates the optical conductivity for MLG on {\hbn} within the framework of model (3) in the THz regime.

\begin{acknowledgments}
This research was funded by EPSRC grant EP/L013010/1 (MM-K), ERC project DM-321031 (DSLA), and Nordita (DSLA).
\end{acknowledgments}


\begin{thebibliography}{99}

\bibitem{novoselov_pnas_2005} K.~S.~Novoselov, D.~Jiang, F.~Schedin, T.~J.~Booth, V.~V.~Khotkevich, S.~V.~Morozov, and A.~K.~Geim, Proc.~Natl.~Acad.~Sci.~USA {\bf 102,} 10451 (2005). 

\bibitem{geim_nature_2013} A.~K.~Geim and I.~V.~Grigorieva, Nature {\bf 499,} 419 (2013).

\bibitem{abergel_advphys_2010} D.~S.~L.~Abergel, V.~Apalkov, J.~Berashevich, K.~Ziegler, and T.~Chakraborty, Adv.~Phys.~{\bf 59,} 261 (2010).

\bibitem{watanabe_natmater_2004} K.~Watanabe, T.~Taniguchi, and H.~Kanda, Nature Mater.~{\bf 3,} 404 (2004).

\bibitem{dean_natnano_2010} C.~R.~Dean, A.~F.~Young, I.~Meric, C.~Lee, L.~Wang, S.~Sorgenfrei, K.~Watanabe, T.~Taniguchi, P.~Kim, K.~L.~Shepard, and J.~Hone, Nature Nanotech.~{\bf 5,} 722 (2010).

\bibitem{haigh_natmater_2012} S.~J.~Haigh, A.~Gholinia, R.~Jalil, S.~Romani, L.~Britnell, D.~C.~Elias, K.~S.~Novoselov, L.~A.~Ponomarenko, A.~K.~Geim, and R.~Gorbachev, Nature Mater.~{\bf 11,} 764 (2012).

\bibitem{xue_natmater_2011} J.~Xue, J.~Sanchez-Yamagishi, D.~Bulmash, P.~Jacquod, A.~Deshpande, K.~Watanabe, T.~Taniguchi, P.~Jarillo-Herrero, and B.~J. LeRoy, Nature Mater.~{\bf 10,} 282 (2011).

\bibitem{yankowitz_natphys_2012} M.~Yankowitz, J.~Xue, D.~Cormode, J.~D.~Sanchez-Yamagishi, K.~Watanabe, T.~Taniguchi, P.~Jarillo-Herrero, P.~Jacquod, and B.~J.~LeRoy, Nature Phys.~{\bf 8,} 382 (2012).

\bibitem{ponomarenko_nature_2013} L.~A.~Ponomarenko, R.~V.~Gorbachev, G.~L.~Yu, D.~C.~Elias, R.~Jalil, A.~A.~Patel, A.~Mishchenko, A.~S.~Mayorov, C.~R.~Woods, J.~R.~Wallbank, M.~Mucha-Kruczynski, B.~A.~Piot, M.~Potemski, I.~V.~Grigorieva, K.~S.~Novoselov, F.~Guinea, V.~I.~Fal’ko, and A.~K.~Geim, Nature {\bf 497,} 594 (2013).

\bibitem{dean_nature_2013} C.~R.~Dean, L.~Wang, P.~Maher, C.~Forsythe, F.~Ghahari, Y.~Gao, J.~Katoch, M.~Ishigami, P.~Moon, M.~Koshino, T.~Taniguchi, K.~Watanabe, K.~L.~Shepard, J.~Hone, and P.~Kim, Nature {\bf 497,} 598 (2013).

\bibitem{kindermann_prb_2012} M.~Kindermann, B.~Uchoa, and D.~L.~Miller, Phys.~Rev.~B {\bf 86,} 115415 (2012).

\bibitem{wallbank_prb_2013} J.~R.~Wallbank, A.~A.~Patel, M.~Mucha-Kruczynski, A.~K.~Geim, and V.~I.~Fal'ko, Phys.~Rev.~B {\bf 87,} 245408 (2013).

\bibitem{wallbank_adp_2015} J.~R.~Wallbank, M.~Mucha-Kruczynski, X.~Chen, and V.~I.~Fal'ko, Ann.~Phys.~(Berlin) {\bf 527,} 359 (2015).

\bibitem{sachs_prb_2011} B.~Sachs, T.~O.~Wehling, M.~I.~Katsnelson, and A.~I.~Lichtenstein, Phys.~Rev.~B {\bf 84,} 195414 (2011).

\bibitem{song_prl_2013} J.~C.~W.~Song, A.~V.~Shytov, and L.~S.~Levitov, Phys.~Rev.~Lett.~{\bf 111,} 266801 (2013).

\bibitem{bokdam_prb_2014} M.~Bokdam, T.~Amlaki, G.~Brocks, and P.~J.~Kelly, Phys.~Rev.~B {\bf 89,} 201404 (2014).

\bibitem{jung_prb_2014} J.~Jung, A.~Raoux, Z.~Qiao, and A.~H.~MacDonald, Phys.~Rev.~B {\bf 89,} 205414 (2014).

\bibitem{jung_natcomm_2015} J.~Jung,  A.~DaSilva, A.~H.~MacDonald, S.~Adam, Nature Commun.~{\bf 6,} 6308 (2015).

\bibitem{dasilva_prb_2015} A.~M.~DaSilva, J.~Jung, S.~Adam, and A.~H.~MacDonald, Phys.~Rev.~B {\bf 91,} 245422 (2015).

\bibitem{wijk_prl_2014} M.~M.~van Wijk, A.~Schuring, M.~I.~Katsnelson, and A.~Fasolino, Phys.~Rev.~Lett.~{\bf 113,} 135504 (2014).

\bibitem{moon_prb_2014a} P.~Moon and M.~Koshino, Phys.~Rev.~B {\bf 90,} 155406 (2014).

\bibitem{san-jose_prb_2014} P.~San-Jose, A.~Gutierrez-Rubio, M.~Sturla, and F.~Guinea, Phys.~Rev.~B {\bf 90,} 115152 (2014).

\bibitem{woods_natphys_2014} C.~R.~Woods, L.~Britnell, A.~Eckmann, R.~S.~Ma, J.~C.~Lu, H.~M.~Guo, X.~Lin, G.~L.~Yu, Y.~Cao, R.~V.~Gorbachev, A.~V.~Kretinin, J.~Park, L.~A.~Ponomarenko, M.~I.~Katsnelson, Yu.~N.~Gornostyrev, K.~Watanabe, T.~Taniguchi, C.~Casiraghi, H-J.~Gao, A.~K.~Geim, and K.~S.~Novoselov, Nature Phys.~{\bf 10,} 451 (2014).

\bibitem{shi_natphys_2014} Z.~Shi, C.~Jin, W.~Yang, L.~Ju, J.~Horng, X.~Lu, H.~A.~Bechtel, M.~C.~Martin, D.~Fu, J.~Wu, K.~Watanabe, T.~Taniguchi, Y.~Zhang, X.~Bai, E.~Wang, G.~Zhang, and F.~Wang, Nature Phys.~{\bf 10,} 743 (2014).

\bibitem{yu_natphys_2014} G.~L.~Yu, R.~V.~Gorbachev, J.~S.~Tu, A.~V.~Kretinin, Y.~Cao, R.~Jalil, F.~Withers, L.~A.~Ponomarenko, B.~A.~Piot, M.~Potemski, D.~C.~Elias, X.~Chen, K.~Watanabe, T.~Taniguchi, I.~V.~Grigorieva, K.~S.~Novoselov, V.~I.~Fal'ko, A.~K.~Geim, and A.~Mishchenko, Nature Phys.~{\bf 10,} 525 (2014).

\bibitem{gorbachev_science_2014} R.~V.~Gorbachev, J.~C.~W.~Song, G.~L.~Yu, A.~V.~Kretinin, F.~Withers, Y.~Cao, A.~Mishchenko, I.~V.~Grigorieva, K.~S.~Novoselov, L.~S.~Levitov, and A.~K.~Geim, Science {\bf 346,} 448 (2014).

\bibitem{song_arxiv_2014} J.~C.~W.~Song, P.~Samutpraphoot, and L.~S.~Levitov, arXiv:1404.4019 (2014).

\bibitem{abergel_njp_2013} D.~S.~L.~Abergel, J.~R.~Wallbank, X.~Chen, M.~Mucha-Kruczynski, and V.~I.~Fal'ko, New J.~Phys.~{\bf 15,} 123009 (2013).

\bibitem{mucha-kruczynski_prb_2013} M.~Mucha-Kruczynski, J.~R.~Wallbank, and V.~I.~Fal'ko, Phys.~Rev.~B {\bf 88,} 205418 (2013). 

\bibitem{jiang_prl_2007} Z.~Jiang, E.~A.~Henriksen, L.~C.~Tung, Y.-J.~Wang, M.~E.~Schwartz, M.~Y.~Han, P.~Kim, and H.~L.~Stormer, Phys.~Rev.~Lett.~{\bf 98,} 197403 (2007).

\bibitem{kuzmenko_prb_2009} A.~B.~Kuzmenko, I.~Crassee, D.~van der Marel, P.~Blake, and K.~S.~Novoselov, Phys.~Rev.~B {\bf 80,} 165406 (2009).

\bibitem{mccann_prl_2006} E.~McCann and V.~I.~Fal'ko, Phys.~Rev.~Lett.~{\bf 96,} 086805 (2006).

\bibitem{mccann_prb_2006} E.~McCann, Phys.~Rev.~B {\bf 74,} 161403 (2006).

\bibitem{house_book_2004} J.~E.~House, \emph{Fundamentals of Quantum Chemistry} (Academic Press, 2003).

\bibitem{footnote_couplings} Some idea about the influence of {\hbn} on the top graphene layer in BLG can be obtained by comparison to the Slonczewski-Weiss-McClure model developed for graphite \cite{dresselhaus_advphys_2002}. Within this model, the direct coupling between the next nearest (carbon) layers is at least an order of magnitude smaller than the direct interlayer coupling.

\bibitem{dresselhaus_advphys_2002} M.~S.~Dresselhaus and G.~Dresselhaus, Adv.~Phys.~{\bf 51,} 1 (2002).

\bibitem{san-jose_prb_2014a} P.~San-Jose, A.~Gutierrez-Rubio, M.~Sturla, and F.~Guinea, Phys.~Rev.~B {\bf 90,} 075428 (2014).

\bibitem{neto_rmp_2009} A.~H.~Castro Neto, F.~Guinea, N.~M.~R.~Peres, K.~S.~Novoselov, and A.~K.~Geim, Rev.~Mod.~Phys.~{\bf 81,} 109 (2009).

\bibitem{gusynin_prb_2006} V.~P.~Gusynin and S.~G.~Sharapov, Phys.~Rev.~B {\bf 73}, 245411 (2006).

\bibitem{abergel_prb_2007} D.~S.~L.~Abergel and V.~I.~Fal’ko, Phys.~Rev.~B {\bf 75,} 155430 (2007).

\bibitem{nicol_prb_2008} E.~J.~Nicol and J.~P.~Carbotte, Phys.~Rev.~B {\bf 77,} 155409 (2008).

\bibitem{zhang_nature_2009} Y.~Zhang, T.-T.~Tang, C.~Girit, Z.~Hao, M.~C.~Martin, A.~Zettl, M.~F.~Crommie, Y.~R.~Shen, and F.~Wang, Nature {\bf 459,} 820 (2009).

\bibitem{varlet_prl_2014} A.~Varlet, D.~Bischoff, P.~Simonet, K.~Watanabe, T.~Taniguchi, T.~Ihn, K.~Ensslin, M.~Mucha-Kruczynski, and V.~I.~Fal’ko, Phys.~Rev.~Lett.~{\bf 113,} 116602 (2014).

\bibitem{footnote_on_low_frequency_behaviour} The divergence of the optical absorption at $\omega\rightarrow 0$ visible in our plots is an artifact resulting from intrinsic imperfections of numerical integration over the sBZ and the finite density of states at $\epsilon=0$ for BLG.

\bibitem{dasilva_arxiv_2015} A.~M.~DaSilva, J.~Jung, S.~Adam, and A.~H.~MacDonald, arXiv:1507.01834 (2015).





\end{thebibliography}
\end{document}